# RA: A machine based rational agent
# Part 1

G. Pantelis

May 21, 2024

**Abstract**

*RA is a software package that couples machine learning with formal reasoning in an attempt to find the laws that generate the empirical data that it has been given access to. A brief outline of RA in its initial stage of development is presented. Particular emphasis is given to current design strategies that aim to endow RA with the ability to construct its own conjectures of which it constructs proofs.*

## 1 Introduction

RA (Rational Agent) is a software package that attempts to find the laws of nature through observations of real-world phenomena. It acquires knowledge by a process of machine learning coupled with formal reasoning.

RA's reasoning is based on a language of computation expressed in the form of programs. In particular it reasons through the formal system PECR (Program Extension Construction Rules) [1]. As such it is not intended that RA will ever become an AI under the more popular conception of an entity that can converse with humans through natural language nor seek knowledge already accumulated by humans. RA's primary objective is to acquire an understanding of the universe from the perspective of the scientific method under the hypothesis that the universe it perceives is computable.

In its current state of development RA is not yet fully autonomous and relies on some human interaction. It is cutoff from the real-world in the sense that it has no sensory apparatus to directly perceive it. Although highly speculative at this stage, we may anticipate the eventual acquisition of such apparatus. These could include telescopic and microscopic vision and other forms of electromagnetic and particle detection. Endowed with such sensory apparatus it is envisaged that RA will acquire the ability to directly observe the real-world and collect its own data from it.

For the foreseeable future RA will rely on a human to supply empirical data. It is within the confines of this data that constitutes the extent of RA's perception of any external world.

## 2 The scientific method

RA is designed to acquire an understanding of the universe by constructing models of it. Working under the hypothesis of a computable universe, computer models of real-world phenomena are built upon laws that are the end product of observations of the real-world through empirical data coupled with rigorous program constructions based on *computability logic* as implemented by the formal system PECR.

The general structure and properties of the main core of any computer model of a real-world application are outlined in Sections 9.7 and 9.8 of [1]. While these computer models can serve as simulation tools they should be regarded as the ultimate expression of the laws that govern real-world phenomena. This opposes the traditional approach of representing the laws of nature by mathematical equations.

The emphasis here is on RA's ability to search for patterns in empirical data and apply formal reasoning to identify the governing laws that generated the data. This should be distinguished from other approaches of model building that rely solely on the direct application of machine learning methods to empirical data. Without intending to diminish some impressive results that can be obtained from the latter, such approaches follow in a spirit that can be associated by analogy with traditional methods of curve fitting.

The formal system PECR coupled with actions of machine learning has a stronger association with the scientific method than the formal procedures of mathematics. In our constructive approach to computability we concede that any application of the formal system PECR may contain valid conjectures that have no proofs that can be derived from the axioms of that application. We make this concession not on the basis of an incompleteness theorem but rather upon the acceptance that there is no general formal procedure that will guarantee in advance that we possess an exhaustive collection of axioms for a given application.

The construction of axioms has largely been a human task that is initiated by some kind of intuitive process, a process that is still not well understood. As a consequence we must acknowledge that in the exploration of any application the list of axioms should ultimately be regarded as provisional. Throughout we will regard axioms to be synonymous with postulates as they are used in the sciences.

An application is denoted by the expression

$S$[pname mach]

where $S$ represents an alphanumeric label that identifies the specific application, pname is the list of names of each *atomic program* of the application and mach is a list of parameters that constrain certain computational operations so that there are no excessive demands made on the finite resources of the machine, M, within which the application is explored (Section 1.5 of [1]). These constraints include bounds on the length of lists of objects of a specific type such as program lists, premise program lists, I/O lists of programs, etc. The list of machine parameter constraints mach can be thought of as defining the depth in which a rational agent chooses to explore an application.

It is within this context that explorations of any application will involve an interplay between formal methods and experimental computation. The overall process will encompass the actions listed in the following table.

| Action | Description |
| --- | --- |
| 1 | Conjecture |
| 2 | Soundness checking |
| 3 | Theorem proving |

Actions 1-2 involve experimental computation while Action 3 is based on formal procedures. Actions 1-3 are run in parallel. The whole process is initiated by finding conjectures that then undergo checks for soundness. Of the surviving conjectures RA will then seek proofs.

A program, p, type[p prgm[n]], can be expressed as an ordered list

p=[p[1] p[2] ... p[n]]

where each p[i], i=1,2,...n, is a fundamental functional program referred to as an *atomic program* (AP). A conjecture comes in the form of an *irreducible extended program* (IEP). An IEP is a program of the form

r=conc[p c]
type[c iexp[p]]

where p is the premise program list and c is the conclusion program, referred to as an *irreducible program extension* (IPE) of the program p.

**In general, a program c is a *program extension* (PE) of a program p and assigned the type ext[p] provided that whenever the program p is computable for a value assignment of its primary input list, pil[p], then the *extended program* (EP), defined by r=conc[p c], type[c ext[p]], is also computable for the same value assigned input. An IPE c of p is a PE of p such that there does not exist a strict sublist q of p such that c is a PE of q.**

An outline of an analogy that exists between natural deduction in proof theory and PEs in PECR can be found in Section 4.9 of [1]. PEs in PECR are always APs. The formal definition of a PE is based on three conditions, the first two being structural and the third based on valuations (Section 4.2 of [1]). Embedded within the first structural condition are the I/O dependency conditions that must be satisfied by any program list (Section 3.2 of [1]).

The list of known IEPs is partitioned by the list

s=[ax th ud]

where ax is the list of axioms, th the list of theorems and ud is the list of IEPs for which no proof is known. Actions 1-3 are temporarily halted whenever (i) a new conjecture is introduced or (ii) a conjecture fails the soundness check or (iii) a proof of a conjecture is found. The lists ax, th and ud are then modified accordingly and Actions 1-3 are restarted with the updated lists ax, th and ud.

Actions 1-3 will be one of an ongoing iteration of revision and feedback. In this way we can identify the overall process as capturing the essential features of the scientific method. Machine learning plays an important role in the overall process. Some specialist in the area of AI might argue that rule based systems will become obsolete and that all inferences can be made under the umbrella of machine learning alone. From the perspective of any application machine learning could identify IEPs of an application without making any distinction where an IEP belongs in the partition s=[ax th ud].

An alternative approach is to couple machine learning with formal reasoning as a way of isolating the fundamental laws (axioms) of a system with the understanding that there exists a formal system based on these laws. This is the preferred approach for those who wish to isolate the fundamental laws of a system from laws that can be derived from them.

## 3 Conjectures

In the current state of affairs the development of methods for automating Actions 2-3 are the least challenging. Identifying new conjectures in the form of IEPs by some automated procedure is still in a highly evolving state and largely occupies the bulk of the work in the current development of RA. There are two procedures involved in the conjecturing process, (i) valuations based on empirical data and (ii) conditions of structure. Conjectures must satisfy certain structural rules that are application independent.

In the previous section we used the notation $S$[pname mach] to represent an application. This differs from the notation used in [1] where an application includes the dependence on a list of axioms. In Chapter 9 of [2] we discussed how an application $S$[pname mach] can exist in any number of states, where each state, $S^{[k]}$[pname ax$^{[k]}$ mach], for some integer k, le[1 k] le[k ms], is associated with a distinct partition of the list of IEPs

$$s^{[k]}=[ax^{[k]} \ th^{[k]} \ ud^{[k]}]$$

Here s$^{[k]}$ represents a distinct partition of all IEPs that exist for the application $S$[pname mach]. Note that we are dealing with a finite system so we can expect that there will exist a finite, albeit possibly large, number, ms, of states. A desirable state $S^{[k]}$[pname ax$^{[k]}$ mach] is one in which the lists ax$^{[k]}$ and ud$^{[k]}$ are of minimal size. A state in which ud$^{[k]}$ is the empty list is said to be *complete*, otherwise the state is said to be *incomplete*.

The difficulty here is that there is no known general formal procedure from which we can determine in advance an optimal collection of axioms for an application. One way to address this issue is to initiate the whole process not by looking for axioms but rather by constructing conjectures. Identifying how a collection of conjectures can be partitioned into axioms, theorems and underivables will be an outcome of the iteration of the parallel Actions 1-3.

Thus the first task is to find a procedure for identifying conjectures that will be more reliable than the intuitive approach of humans. RA regards conjectures as provisional IEPs that are initially identified as elements of the list ud. Each new conjecture is subjected to checks for soundness. When proofs of conjectures are found they become elements of the list th. An IEP for which there is no known proof but is employed in the proof of at least one theorem is regarded as an axiom and becomes an element of the list ax.

Many IEPs exhibit structural similarities when viewed through the decompositions of their I/O matrices (Section 10.3 of [1]). In particular IEPs whose *binary I/O binding matrix templates* are identical are said to be structurally equivalent. Exploring patterns in the configurations of the I/O binding matrices of programs reveal the essential structural features that ultimately lead to the identification of conjectures.

The process of identifying conjectures in RA is still in its infancy and largely remains a work in progress. At this stage of its development RA constructs conjectures by first searching through possible configurations of binary I/O binding matrix templates of programs that are most relevant to the application under consideration.

The conjecturing process is carried out with strict adherence to the structural conditions of PEs as outlined in Section 4.2 of [1]. Embedded within the first of the two structural conditions of a PE are the I/O dependency conditions that are crucial to any program list. The PE structural conditions are based on straight forward computations that significantly reduce the search for configurations of binary I/O binding matrix templates that can exist in any application.

Because of the rapid evolution of this component of RA we will refrain from supplying a detailed outline of the whole process in its current state of development. More will be said on the importance of I/O binding matrices in the conjecturing process in Part 2 of this article where the results of a preliminary test of RA are presented [3].

## 4 Soundness

In the scientific method a postulate is regarded as valid only up until such time that it is found to be inconsistent with observation. Following in this spirit we have at our disposal a procedure for detecting invalid IEPs by experimental computation. This involves checking for computability by executing a program for value assignments of its primary input list obtained from a domain of all possible valid value assignments of the program.

A necessary condition for the validity of an IEP is that it first be an EP as outlined in Section 4.2 of [1]. PE integrity tests are based on two structural conditions followed by a condition based on value assignments. The structural integrity tests for PEs are performed in Action 1 and act as an initial filter for the search of valid binary I/O binding matrix templates.

In PECR soundness is defined by the valuations component of the PE integrity conditions. In RA soundness checking is performed as follows ::

- The premise program, p, is executed for value assignments of its primary input list, pil[p]. If the premise program, p, of an EP r=conc[p c], type[c ext[p]], is computable for a given value assignment of its primary input list then it must follow that r is also computable for the same value assigned input. Otherwise r=conc[p c] is not an EP.

The soundness of a statement of falsity, type[p false], can be checked in a slightly different way ::

- The premise program, p, is executed for value assignments of its primary input list, pil[p]. If it is found that there exists a value assignment of the primary input list such that p is computable then the statement of falsity fails the soundness check.

It can be shown that soundness of a statement of falsity can be incorporated into the PE soundness conditions in an expanded context of HOPs (Section 4.10 of [1]).

Note that the above tests for soundness only check that the conjecture is an EP. Conjectures that come in the form of EPs but are not IEPs can be eliminated through RA's automated theorem prover that has the ability to identify superfluous statements in the premise program. It does this by the process of *connection list reduction* as outlined in Section 10.7 of [1].

All IEPs start off as conjectures that are regarded as provisional and subject to ongoing checks for soundness by Action 2. In the experimental tests for soundness, value assignments of the primary input list, pil[p], of an EP, r=conc[p c], type[c ext[p]], are selected from a domain of all possible valid input values of the program p by a combined systematic and random process. The domain of valid value assignments of programs is inferred from the empirical data associated with the application. As a consequence of the combination of Actions 1 and 2 conjectures become the end products of checks for structural integrity and consistency with observation.

The soundness checking procedure can be regarded as a search for counter examples to an assumption of soundness. Such tests are conducted on all conjectures and are run continuously and parallel to Actions 1 and 3. Confidence building of a conjecture as an IEP proceeds as it survives the ongoing checks for soundness through experimental computations.

While the soundness checking methods described here are largely based on confidence building they have been found to be remarkably effective in identifying invalid IEPs. Actions 1-3 are temporarily halted when a conjecture is found to violate the soundness condition. The invalid conjecture is discarded along with all proofs of elements of th that depend on the invalid conjecture.

## 5 Theorem proving

Conjectures are fed into RA's theorem prover where proofs of the conjectures are sought. Proofs in PECR differ from traditional proofs of mathematics in that they are based on the rigorous constructions of programs employing a version of computability logic as implemented by the formal system PECR. At each step of a proof of a conjecture all possible PEs of the current proof program are generated. This involves extracting all of the sublists from the current proof program that are I/O equivalent to premises of known IEPs (Section 4.5 of [1]).

RA has the additional task of determining what can be regarded as the fundamental laws of an application that distinguish them from the laws that can be derived from them. This essentially involves the construction of a suitable state, $S^{[k]}[\text{pname } ax^{[k]} \text{ mach}]$, for some k, le[1 k] le[k ms], that is associated with a unique partition of the list of IEPs.

Initially an IEP will be regarded as underivable, i.e. an element of the list ud of the partition s=[ax th ud]. When a proof of an IEP is found RA will regard it as an element of th. If an element of ud is found to be employed in the proof of at least one theorem it is reassigned as an element of ax.

In practice the process of choosing an optimal partition of the list of IEPs is not straight forward. There does not appear to be any general formal procedure that can identify in advance the properties of fundamental laws that can distinguish them from laws that are derivable from them. It is often speculated that the laws of nature are governed by simple computational rules suggesting that a primary property of a fundamental law, expressed as a program, will be one of minimal length. However, the minimal program length of elements of ax does not guarantee that the length of the list ax is minimal. Therefore there is an interplay between the complexity of each fundamental law versus the complexity of the list of fundamental laws that govern an application.

In light of this RA makes no initial assumptions on whether a conjecture is an axiom or theorem. Suppose the dependence of a proof of an IEP k can be traced back to an IEP j such that IEP j has a proof whose dependence can be traced back to IEP k. When this occurs RA makes a decision on whether to seek another proof of IEP k that is independent of IEP j or seek another proof of IEP j that is independent of IEP k. RA will continue this process of choice in eliminating and reconstructing proofs until it is satisfied that it has found a proof of an IEP that is independent with respect to proofs of all known IEPs. Once an independent proof of an IEP is found Actions 1-3 are temporarily halted and the IEP becomes an element of the list th.

Each theorem comes with a *theorem connection list*. A theorem connection list tcl[k] is a sublist of the list of labels of all known IEPs and contains the labels of the IEPs that were explicitly employed in the proof of IEP k. Theorem connection lists exclude the IOT axiom and the substitution rule. Elements of ax and ud have empty theorem connection lists.

Through the theorem connection lists we can trace back all dependencies of a proof. The algorithm can be written as follows.

*Tracing back all dependencies of the proof of* IEP k

```
b:=tcl[k]
do
    c:=b
    do i=1 to len[b]
        c:=unique[conclst[c tcl[b[i]]]]
    end do
    if b=c exit
    b:=c
end do
```

Upon convergence the list b will contain the labels of all IEPs that can be traced back as dependencies of the proof of IEP k. The proof of IEP k will be regarded as independent if k is not an element of the final output of the list b.

It is possible that an independent proof is found for an IEP that is currently labeled as an axiom. This can occur after a new conjecture is added to the list of IEPs by Action 1. When this occurs the IEP that is currently labeled as an axiom is relabeled as a theorem.

The automated theorem prover of RA can also be used as a stand alone package. One simply creates a list of axioms and conjectures that are then directly fed into RA's theorem prover where proofs of the conjectures are generated. Archived proofs of theorems associated with various applications that were originally constructed interactively using the proof assistant software VPC have been replaced by those generated by RA [1]-[2].

Most of the proofs generated by RA are very similar to those that were constructed interactively with VPC. However, there are proofs where RA exhibits some kind of unexpected strategy that appears at first glance to be alien to human generated proofs. This phenomena is not new and has been observed in many AI gaming applications. Examples of this phenomena are discussed in Part 2 of this article where the results of a preliminary test of RA are presented [3].